\documentclass{ifacconf}

\usepackage{graphicx}      
\usepackage{natbib}   
\usepackage{url}

\usepackage{amsmath,amssymb,amsfonts}
\usepackage{algorithmic}
\usepackage{graphicx}
\usepackage{textcomp}
\usepackage{bm}
\usepackage{subcaption}

\usepackage{siunitx}

\usepackage{soul}

\usepackage{empheq}
\let\cite\citep

\DeclareMathAlphabet\mathbfcal{OMS}{cmsy}{b}{n}

\newcommand*{\genbf}[1]{\ifmmode\mathbf{#1}\else\textbf{#1}\fi}
\newcommand{\norm}[1]{\left\lVert#1\right\rVert}

\newcommand{\yrefb}[0]{\mathbf{y_{ref}}}
\newcommand{\Yrefb}[0]{\mathbf{Y_{ref}}}
\newcommand{\Db}[0]{\mathbf{D}}

\newcommand{\bs}[0]{\mathbf{s}}
\newcommand{\bd}[0]{\mathbf{d}}
\newcommand{\bp}[0]{\mathbf{p}}

\newcommand{\Emme}[0]{\boldsymbol{\mathcal{M}}}

\newcommand{\xb}[0]{\mathbf{x}}

\newcommand{\ub}[0]{\mathbf{u}}

\newcommand{\db}[0]{\mathbf{d}}

\newcommand{\etab}[0]{\boldsymbol{\eta}}

\usepackage{tikz}
\usetikzlibrary{arrows}

\newtheorem{remark}{Remark}

\newtheorem{problem}{Problem}

\begin{document}
\begin{frontmatter}


\title{Resilient AFE Drive Control using Neural Networks with Tracking Guarantees\thanksref{footnoteinfo}} 

\thanks[footnoteinfo]{This research has been supported by the Swiss National Science Foundation under the NCCR Automation (grant agreement 51NF40\_180545) and the NECON project (grant number 200021-219431).}%

\author[First]{Nicolas Kirsch} 
\author[Second]{Catalin Arghir} 
\author[Second]{Silvia Mastellone}
\author[First]{Giancarlo Ferrari-Trecate}

\address[First]{École Polytechnique Fédérale de Lausanne, DECODE lab, Lausanne, Switzerland (e-mail:\{nicolas.kirsch, giancarlo.ferraritrecate\}@epfl.ch)}
\address[Second]{FHNW University of Applied Sciences and Arts Northwestern Switzerland, Windisch, Switzerland (email:silvia.mastellone@fhnw.ch,carghir@ethz.ch)}
\begin{abstract}                
Industrial installations across several sectors have seen a dramatic increase in
 productivity, accuracy and efficiency over the last decade due to expanded utilization of medium voltage, variable speed  power electronic converters to drive their processes.
Specifically, active front-end (AFE) drives have become  popular due to their ability to deliver power while maintaining safe electrical setpoints. However, under abnormal grid conditions such as phase loss, conventional AFE control may fail to enforce safety constraints, potentially leading to drive shutdown and significant financial losses.

In this work, we propose using reference-tracking Performance Boosting (rPB) to improve the resilience of standard AFE control to faults. This neural-network control framework provides a principled way to optimize transient performance while preserving the steady-state tracking properties of AFE-based drives. By carefully shaping the input signals to the rPB controller, we ensure that it activates only during grid faults, leaving nominal operation unaffected. Simulation results show that the proposed approach successfully maintains the DC bus voltage and the grid current within safe limits during single-phase loss events.

\end{abstract}


\end{frontmatter}

\section{Introduction}

Modern power grids are increasingly adopting active front-end (AFE) converters serving not only in interfacing renewable generation, such as wind and hydro, but also in flexible industrial loads including large compressors, heat pumps, and pumped-hydro drives \citep{hokayem2024control}. These systems may require bidirectional power flow, operating alternately as energy consumers and sources, and thus form the cornerstone of a new class of grid-interactive electrical assets. Apart from handling the AC current and regulating the DC bus, AFEs may enable ancillary services such as reactive power compensation and inertia emulation, even in traditional motor drive applications \citep{arghir2025transferring}.

These functions are typically implemented through a standard cascaded PI-based control architecture, including an inner loop for current regulation, and an outer DC-bus voltage controller, delivering high performance under nominal grid conditions. However, perturbed grid states such as voltage sags, asymmetrical faults, or weak-grid operation significantly degrade converter performance \citep{10603443, he2024cross, stanojev2025grid}. Under these disturbances, the AFE struggles to keep its controlled variables within constraints, particularly the DC-bus voltage, which may experience severe under-voltage. Such deviations frequently trigger protective shutdowns, interrupting industrial processes and leading to significant productivity losses \citep{10100631, eslahi2023resiliency}. 
%
%
%
Typical fault ride-through strategies rely on modifying the current reference, either through dynamic limiters \citep{TANG2025111407} or mode-switching logic that adapt the reference upon fault detection \citep{6584824}. While effective, limiter-based approaches can lead to current distortion during asymmetrical faults, and mode-switching methods are usually limited to short-circuit or deep-sag events \citep{8371651}. Maintaining both continuous operation and safe, bounded dynamics during large electrical disturbances remains an open issue in modern AFE-equipped drives \citep{11154947}.

Within the broader theory of reference tracking for nonlinear systems, several approaches have been considered to improve transient performance. Model Predictive Control (MPC) approaches can track time-varying references while incorporating secondary, potentially nonlinear objectives through constraints or the loss function \citep{mpc_rt}, sometimes with closed loop guarantees \citep{mpc_rt2}. However, the deployment of MPC policies typically requires solving complex optimization problems in real time. This can be computationally impractical when dealing with highly non-linear models and cost functions \citep{rawlings2017model}.

In adaptive control, the controller is adjusted in real-time based on the system's performance \citep{Nguyen2018}, and constraint satisfaction is typically enforced using barrier functions \citep{output_barrier} or safety filters \citep{schneeberger2024advanced}. While these methods are promising, closed-loop guarantees are usually restricted to stability and constraint satisfaction and often apply locally.

Neural networks have also been studied to improve reference tracking, but mainly as observers \citep{PARK2017353, gao2019long} or to optimize tracking metrics, often without providing guarantees \citep{zhang2022neural}. 
The authors of \citep{kirsch2025boostingtransientperformancereference} introduced a neural network controller with reference tracking guarantees. This neural controller works as a plug-in supporting pre-existing reference tracking controllers during transient while not interfering   in steady state. Most importantly, \citep{kirsch2025boostingtransientperformancereference} shows how to parametrize \textit{all and only} reference tracking controllers in terms of a free operator satisfying mild input to output conditions. This allows improving the performance of the controller for any nonlinear objective using unconstrained optimization methods, enhancing  the expressivity and reducing the computational burden. Simulation results have shown promising improvement on simple systems, making this method seemingly valuable for the support of standard active front end control during faults.

\subsection*{Contributions} 

In this paper, we apply the novel neural network–based controller architecture known as rPB to AFE drives. Our goal is to design an outer control loop that maintains stable operation even under grid faults. The rPB framework parametrizes all and only reference-tracking controllers in terms of a free operator, which enables us to optimize nonlinear safety objectives in an unconstrained manner while preserving the tracking properties of the AFE controller. By carefully designing the input signals, we ensure that the rPB controller activates precisely when required, and only then, even in the presence of prolonged or sequential faults.

To demonstrate the effectiveness of the proposed approach, we test the controller on a  scenario of  single-phase  voltage drops of varying amplitudes in the supply grid. The neural network is trained on a realistic drive simulation, and results show that the rPB controller successfully maintains system safety even in the event of a complete loss of one grid phase.

The paper is structured  as follows: Section 
\ref{Sec2} describes the system and the AFE controller.
Section \ref{Sec3} introduces the rPB control approach, formulates the converter control design problem   and details the solution approach. Section 
\ref{Sec4} demonstrates the controller effectiveness in simulation based case studies and finally Section \ref{Sec5} contains concluding remarks and future work.
\subsection*{Notation}
Across the paper we will be adopting the following notation: the set of all sequences $\mathbf{x} = (x_0,x_1,x_2,\ldots)$, where $x_t \in \mathbb{R}^n$, $t\in \mathbb{N}$ , is denoted as $\ell^n$. 
Moreover,  $\mathbf{x}$ belongs to $\ell_p^n \subset \ell^n$ with $p \in \mathbb{N}$ if $\norm{\mathbf{x}}_p = \left(\sum_{t=0}^\infty |x_t|^p\right)^{\frac{1}{p}} < \infty$, where $|\cdot|$ denotes any vector norm. 
We say that $\xb \in \ell^n_\infty$ if $\operatorname{sup}_{t}|x_t|< \infty$. 
When clear from the context, we omit the superscript $n$ from $\ell^n$ and $\ell^n_p$. 
An operator $\mathbf{A}:\ell^n \rightarrow \ell^m$ is said to be \emph{causal} if $\mathbf{A}(\mathbf{x}) = (A_0(x_0),A_1(x_{1:0}),\ldots,A_t(x_{t:0}),\ldots)$, and $\mathbf{A}$ is said to be $\ell_p$-stable if it is \emph{causal} and $\mathbf{A}(\mathbf{w}) \in \ell_{p}^m$ for all $\mathbf{w} \in \ell_{p}^n$, with $p\in \mathbb{N}\cup\{\infty\}$. Equivalently, we  write $\mathbf{A} \in \mathcal{L}_{p}$. 
We say that an $\mathcal{L}_p$ operator $\mathbf{A}:\mathbf{w}\mapsto \mathbf{u}$  has finite $\mathcal{L}_p$-gain $\gamma(\mathbf{A})>0$ if  $\|\mathbf{u}\|_p\leq \gamma(\mathbf{A})\|\mathbf{w}\|_p$, for all $\mathbf{w}\in\ell_p^n$.

\section{System description}
\label{Sec2}

\begin{figure}[h]
    \centering
    \includegraphics[width=\linewidth]{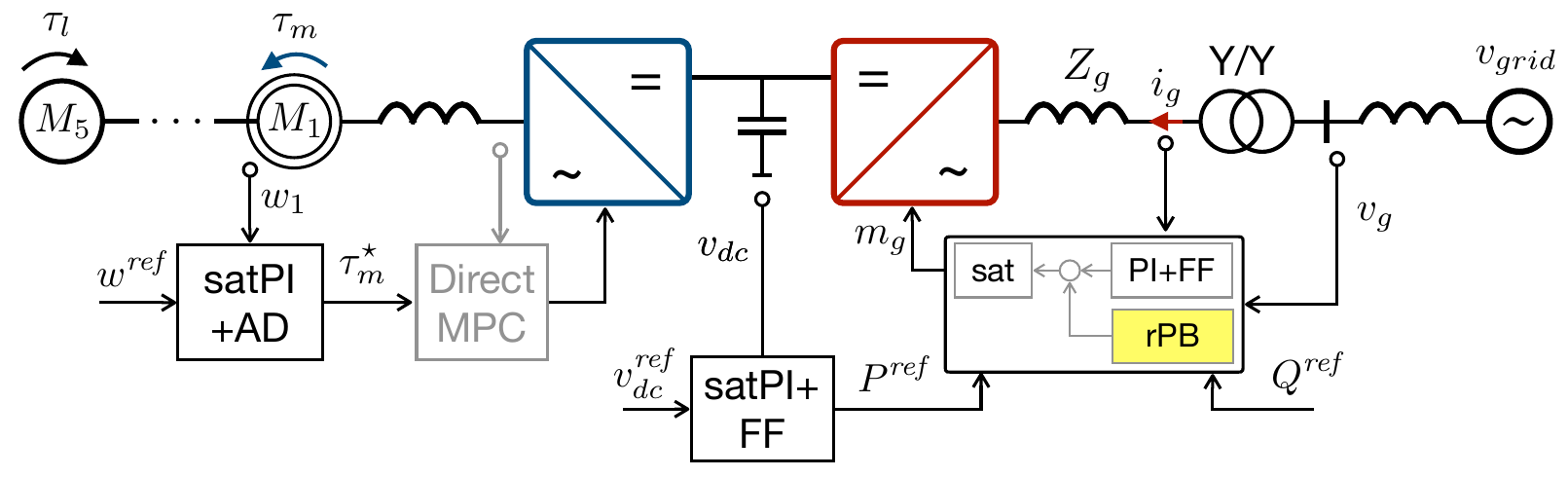}
    \caption{Block diagram of a typical AFE-drive system with a complex driveline shaft. The control includes inner and outer loops implementing anti-windup PI compensators with active damping (AD) and feed-forward (FF) terms where appropriate. In yellow we highlight the proposed rPB control augmentation. 
    }
    \label{fig:dtcdrive}
\end{figure}

We consider a medium-voltage drive system, where the AFE is implemented using a three-level active neutral-point-clamped (ANPC) converter. On the load-side, we consider a three-level ANPC converter connected to an Induction Machine (IM) model whose rotor is part of a 6MW complex driveline shaft model. 
The air-gap torque $\tau_m$ acts on one end of the shaft and the load $\tau_l$ on the other. The motor torque is produced through (an industry-standard) direct MPC which takes $\tau_m^\star$ as reference. This setup is depicted in Fig. \ref{fig:dtcdrive}. 

\subsection{Drive system model}

For our control design purpose, we shall omit the motor and the torque control dynamics and assume that the  torque $\tau_m$ perfectly tracks the reference $\tau_m^\star$. 
We adopt the energy-preserving, average-switch model of the back-to-back drive system used in \citep{chandrasekaran2025reactivepowerflowoptimization, arghir2025transferring}, which we discretise using forward Euler with a step time of $h=2.5*10^{-4}s$, yielding: 
%
%

\begin{subequations}
\label{DTC_FE}
\begin{align}
    w_{t+1} &= (1-\frac{hD}{M})w_t + \frac{h}{M}\tau_{m,t} - \frac{h}{M}\tau_{l,t} 
    \\
    v_{dc,t+1} &= (1-\frac{hG_{dc}}{C_{dc}})v_{dc,t} - \frac{h}{C_{dc}} \frac{\tau_{m,t}}{v_{dc,t}}w_t + \frac{h}{C_{dc}}m_{g,t} ^\top i_{g,t} 
    \\
    i_{g,t+1} &= (I -hL_g^{-1}R_g) i_{g,t} + hL_g^{-1}v_{g,t} -hL_g^{-1}m_{g,t} v_{dc,t}
\end{align}
\end{subequations}

where \( M, D \) are the moment of inertia and viscous damping of the driveshaft, \( w \) its angular velocity, \( \tau_m \) is a control input, \( \tau_l \) is a disturbance, \( v_{dc} \) is the DC-link voltage, \( C_{dc} \) its capacitance and \( G_{dc} \) its parallel conductance. Furthermore, \( m_g \in \mathbb{R}^2 \) is the grid-side converter modulation vector (control input), \( L_g, R_g \in \mathbb{R}^{2\times 2} \) the phase inductance and resistance respectively, with strictly positive entries. The vector \( i_g \in \mathbb{R}^2 \) is the grid current (towards the converter) and \( v_g \in \mathbb{R}^2 \) is the voltage (measured disturbance). All two-dimensional quantities are considered to be power-invariant \( \alpha\beta \)-coordinate representations of three-phase signals. Overall, \eqref{DTC_FE} is a nonlinear model with states $[w,v_{dc},i_g]$, inputs $[\tau_m,m_g]$, and disturbances $[\tau_l,v_g]$.

Given the fast response of the inner motor control loop, we assume that we can directly actuate $\tau_m$ in \eqref{DTC_FE}. Our motivation is based on the fact that most modern MV motor controllers adopt a form of Direct Torque Control (DTC) \citep{Geyer2008,Tiitinen1996}, to track the torque set-point with high bandwidth. 


We adopt the state of the art control structure described in \citep{chandrasekaran2025reactivepowerflowoptimization} and shown in Figure \ref{DTC_FE}. An outer loop PI controller is designed to  track $w^\textit{ref}$ through $\tau_m$, and the two inner loop PI controllers  are designed to track $v_{dc}^\textit{ref}$ and $Q^\textit{ref}$ by actuating $m_g$, where $Q = v_q i_d-v_d i_q \in \mathbb{R}$ is the reactive power from the grid. The reference signals are constant over the time horizon and lie within the admissible operating range of the considered drive.

\section{Control methodology} \label{Sec3}
The standard AFE control strategy ensures accurate tracking and constraint satisfaction under nominal operation, such as steady load torque $\tau_l$ and balanced, constant grid voltage $v_g$. However, when exposed to perturbed conditions, the controller may fail to maintain the DC bus voltage within its allowable operating range, potentially triggering system shutdown.

To address this, our objective is to design an auxiliary outer control loop that supports the existing controller during fault transients, ensuring constraint satisfaction without degrading nominal tracking performance. This additional layer should remain inactive during steady-state operation, thereby preserving the high-quality reference tracking of the base  controller. The reference tracking performance boosting framework presented in the following section provides a suitable option to achieve this balance between fault compensation and nominal performance. 

\subsection{rPB Controllers} \label{rpb control}

The rPB controller is a plug-in neural network controller which boosts the transient performance of a pre-existing tracking controller while preserving its guarantees in steady-state. It does so by learning to solve a Nonlinear Optimal Control problem (NOC). Its objective is not to replace controllers achieving steady-state reference tracking, but to support them during aggressive  transients that might cause extreme voltage deviation in the capacitor  and lead to system shutdown. 


We can make the drive dynamics compatible with the rPB framework by reformulating them as follows. First, we introduce the concatenated sequence $\etab \in \ell^{8}$, which collects the full closed loop states between the system and the PIs. It consists of the three system states, one of which is two-dimensional, along with the three PI integrator states, one of which is also two-dimensional. We do the same for the references, giving $\yrefb = [w^\textit{ref},v_{dc}^{\textit{ref}},Q^\textit{ref}] \in \mathbf{Y_{ref}}$, with $\mathbf{Y_{ref}}\subset\ell^3$ being the set of all references which can be tracked by the base system. 

Within the rPB framework, it is assumed that the system is subject to an additive disturbance signal in $\ell_p$. However, neither the load torque $\tau_l$ nor the grid voltage $v_g$ satisfies this assumption, as neither asymptotically decays, even in the presence of faults. To address this, we decompose the grid voltage into two components: the nominal voltage $\bar{v}_g$ and a perturbation term $\delta v_g$, such that
\begin{equation*}
    v_g = \bar{v}_g + \delta v_g.
\end{equation*}
During grid faults, the supplied voltage deviates from its nominal value, implying $\delta v_g \neq 0$, while otherwise $\delta v_g = 0$. Under the assumption of a finite number of faults with finite duration, the sequence $\boldsymbol{\delta v_g}$ belongs to $\ell_p$.
Since $v_g$ enters the system linearly, we can treat $\boldsymbol{\delta v_g}$ (scaled by the factor $hL_g^{-1}$) as the required $\ell_p$ additive disturbance within the rPB formulation. Finally, we encode the system’s initial condition within this disturbance, yielding the sequence $\mathbf{p} = \bigl( p_0, p_1, \ldots, p_T \bigr) \in \ell^{10}$, where:


\[
p_0 = \begin{bmatrix}\eta_0 \\ 0_{2\times 1}\end{bmatrix},\;
p_t = \begin{bmatrix}0_{8\times 1} \\ h L_g^{-1}\, \delta v_{g,t}\end{bmatrix}
\quad \text{for } t = 1,\ldots,T,
\]

where $T\in \mathbb{R}$ is the horizon. The disturbance which are not $\ell_p$ sequences are also concatenated, giving $\db=[\tau_l,\bar{v_g}]\in \mathbf{D}$, with $\mathbf{D} \subset \ell^3$ the set of acceptable nominal disturbances. 

We also define $\ub \in \ell^m $ as the control action coming from the rPB controller. There is no assumptions as to how $\ub$ influences the base system.  Letting $f:\mathbb{R}^8\times\mathbb{R}^m\times\mathbb{R}^3\times\mathbb{R}^3\rightarrow\mathbb{R}^8$ represent the closed loop dynamics of the base system with controller, we obtain: 
\begin{equation} \label{eq:system_state}
    \eta_{t+1} = f(\eta_t,u_t,y_{ref,t},d_t) + p_t ~~~t= 1,2,\ldots\,,.
\end{equation}

Model \eqref{eq:system_state} captures all components of Figure \ref{fig:dtcdrive} except for the yellow block. In operator form, \eqref{eq:system_state} can be written as

\begin{equation}
\label{eq:operator_form_state}
    \boldsymbol{\eta} = \mathbf{F}(\boldsymbol{\eta}, \ub,\mathbf{y_{ref}},  \bd) + \bp\,,\end{equation}
where $\mathbf{F}:\ell^8\times\ell^3\times\ell^m\times\ell^3\rightarrow\ell^8$ is a strictly causal operator. 

A key assumption of the rPB framework is that the base system asymptotically tracks the references. This means that for all $\db \in \mathbf{D}, \yrefb\in \mathbf{Y_{ref}}$, $\mathbf{p} \in \ell_p$, and for $\ub = 0$, the tracking error asymptotically goes to zero. In our case, the assumption is satisfied as the cascading PI architecture tracks the references under nominal conditions.

\begin{figure}[h]
    \centering
    \includegraphics[width=\linewidth]{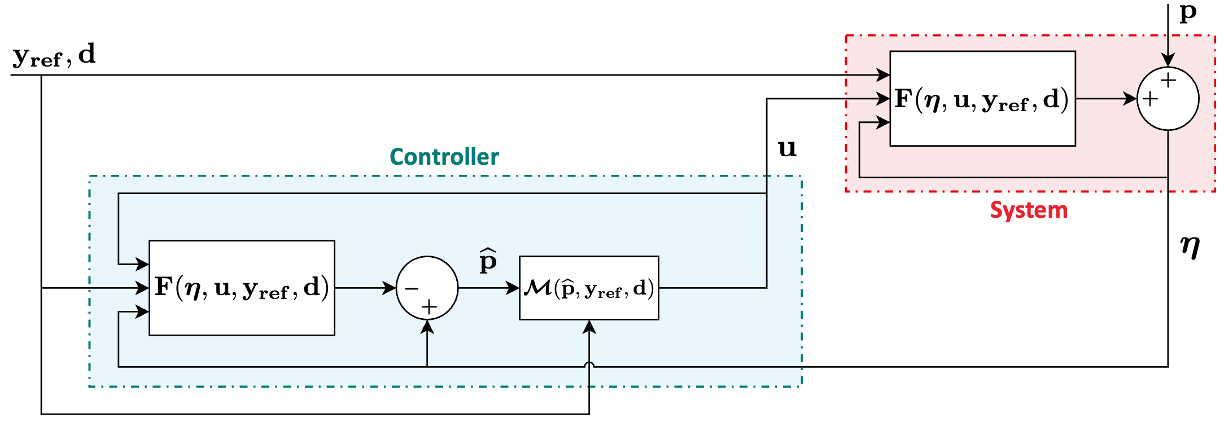}
    \caption{IMC architecture parametrizing all reference tracking controllers in terms of one operator $\Emme$}
    \label{fig:IMC Block}
\end{figure}

In \citep{kirsch2025boostingtransientperformancereference}, it is shown that if this condition is met, then under an Internal Model Control (IMC) architecture, all and only controllers capable to track the same references can be parametrized in terms of an operator $\Emme$ generating $\ell_p$ signals when $\bp \in  \ell_p$. The IMC control architecture, which is shown in Figure \ref{fig:IMC Block}, includes a copy of the system dynamics, which is used for computing the estimate $\widehat \bp = \etab-\mathbf{F}(\etab,\mathbf{u},\yrefb, \bd)$ of the disturbance $\bp$. In this setup, $\ub$ is chosen as: 

\begin{equation}
     \ub = \Emme(\widehat \bp, \yrefb,\bd),
\end{equation}

for a causal operator $\Emme:\ell^8\times \ell^3\times \ell^3 \rightarrow \ell^m$. The intuition behind this control architecture is that given $\widehat\bp \in \ell_p$, the output of the rPB controller will also be an $\ell_p$ signal. It will thus decay to zero when the base system approaches steady state condition, thereby enabling the tracking of the reference. 

The operator $\Emme$ thus needs to generate $\ell_p$ sequences when $ \widehat\bp$ is an $\ell_p$ signal, and irrespectively of $\yrefb$ and $\db$ which are not necessarily in $\ell_p$. To this end, we factorize $\Emme$ as follows:

\begin{equation} \label{factorization}
    \Emme(\widehat \bp, \yrefb,\bd) = \Emme_2(\widehat \bp)*\Emme_\infty(\widehat \bp, \yrefb,\bd),
\end{equation}

with $\Emme_2 \in \mathcal{L}_p$ and $\Emme_\infty(\widehat \bp, \yrefb,\bd) \in \ell_\infty, \,\,  \forall \widehat \bp \in \ell_p, \yrefb \in \Yrefb $ and $\bd \in \Db$. The boundedness of $\Emme_\infty$ ensures that $ \Emme(\widehat \bp, \yrefb,\bd) \in \ell_p$.


As shown in \citep{kirsch2025boostingtransientperformancereference}, searching over operators $\Emme$ factorized as in \eqref{factorization} allows one to improve performance while preserving steady-state guarantees. However, optimizing directly over such $\Emme$ constitutes an infinite-dimensional problem and is therefore intractable. To make the NOC formulation tractable, \citep{kirsch2025boostingtransientperformancereference} proposes restricting the search for $\Emme_2$ to subsets of $\mathcal{L}_2$ that are described by a finite number of free parameters, while still being expressive enough to represent broad families of nonlinear controllers. This can be achieved using several operator models introduced in the literature, such as Recurrent Equilibrium Networks \citep{REN} or Structured State-Space Models \citep{gu2021efficiently,massai2025free}. In these frameworks, $\mathcal{L}_2$ operators are represented as parametrized dynamical systems interconnected with neural networks, enabling rich nonlinear behavior. The operator $\Emme_\infty$ can then be parametrized as a Multilayer Perceptron (MLP) with a bounded output activation function, which also has a finite number of parameters.  

By using the direct parametrization of \cite[Section 5]{REN} for $\Emme_2$ and leveraging the naturally free parametrization of MLPs for $\Emme_\infty$, the rPB optimization problem becomes fully unconstrained. The parameters to optimize over become $\theta \in \mathbb{R}^d$, which includes the parameters of both $\Emme_2$ and $\Emme_\infty$. 

We want to improve performance over a horizon $T$ by solving a finite-horizon NOC, while also ensuring tracking is preserved even when the controller is applied for $t > T$. Considering an empirical average over a set of samples $\{(p_{T:0}^s,d_{T:0}^s)\}_{s=1}^S$ drawn from the distributions $\mathcal{P}_{T:0}$ and $\mathcal{D}_{T:0}$, this NOC problem can be expressed as:\\


\begin{problem}{Finite horizon unconstrained problem}
\label{prob:boosting2}
 \begin{subequations}
    \label{NOC2:cost_and_stab_free}
	\begin{alignat}{3}
	&\min_{\theta \in \mathbb{R}^d}&& \qquad \frac{1}{S} \sum_{s=1}^SL(\eta^s_{T:0},u^s_{T:0}, y^s_{ref,T:0})\label{NOC2:cost}\\
	&\operatorname{s.t.}~~ && \eta^s_t = f_t(\eta^s_{t},u^s_{t},y^s_{ref,t},d^s_t)+ w^s_t\,, \nonumber \\ 
    &~~&&~~\phantom{\eta^s_t = f_t(\eta^s_{t-1:0},u^s_{t-1:0},)\,\,\,} w^s_0 =(x^s_0,v^s_0)\,,\label{Model} \\
	&~~&&u^s_t = \mathcal{M}_t(\theta)(w^s_{t:0},y^s_{ref,t:0},d^s_{t:0})\,,~~\forall t =0,1,\ldots\,, \label{NN}
	\end{alignat}
 \end{subequations}
 \end{problem}
where $\eta_{T:0}^s$ and $u_{T:0}^s$ are the inputs and states obtained when the disturbances $p_{T:0}^s$ and $d_{T:0}^s$ are applied. 
As shown in \citep{kirsch2025boostingtransientperformancereference}, the absence of constraints on $\theta$ allows us to leverage powerful optimization frameworks such as PyTorch \citep{paszke2019pytorch}, using a backpropagation-through-time approach \citep{werbos1990backpropagation} to design the rPB controller efficiently.

The cost $L(\cdots)$ in \eqref{NOC2:cost} is completely free and can account for any behavior that is wished to be enforced. For the drive, as discussed in Section \ref{Sec4}, the function $L$ can be chosen for example to promote constraint satisfaction by using barrier functions. The rPB controller thus provides a way to improve the transient behavior of the system for any desired goal, without hindering the steady-state behavior induced by the base controller.

\subsection{Input signal design} \label{signal}

As described in the previous section, rPB can successfully preserve steady state guarantees because it outputs by construction $\ell_p$ signals when $\ell_p$ sequences are fed to $\Emme_2$. 
However, this intrinsic property also implies that to successfully improve transient performance, the input signals to the rPB controller needs to be carefully designed. If not, the output of the rPB controller could be active when not needed, but more importantly it could already be decaying when it is required. The ideal signal would be non-zero during transient periods and immediately zero when the rPB controller is not needed anymore. Since $\boldsymbol{\delta v_g}$ is nonzero only during faults, these conditions are naturally satisfied by $\mathbf{p}$. Feeding its reconstruction $\mathbf{\widehat p}$ to $\Emme_2$ excites the operator during the faults and not during nominal conditions.  

\begin{figure}[h]
    \centering
    \includegraphics[width=0.9\linewidth]{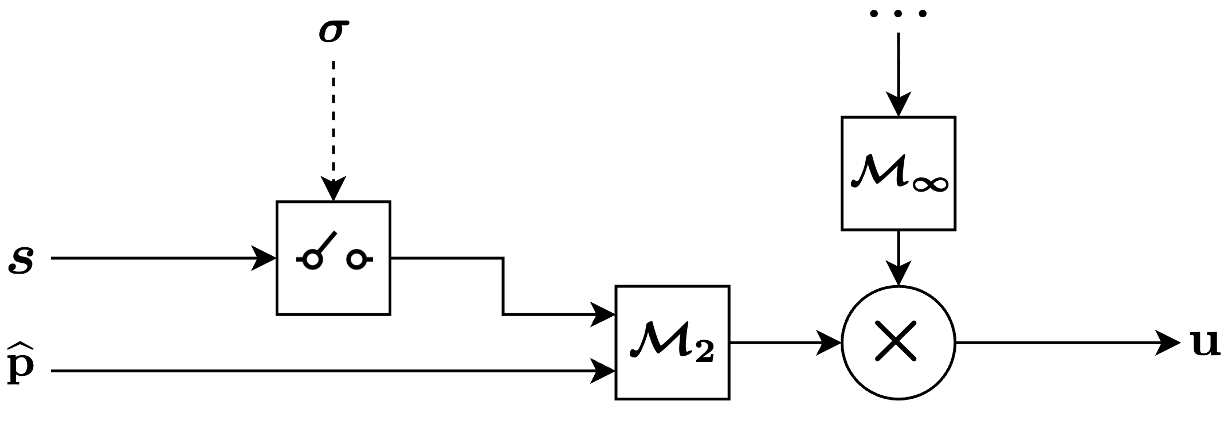}
    \caption{Input signal structure to the operator $\Emme$. The switch logic is dictated by the condition on $\boldsymbol \sigma$. The signals entering $\Emme_\infty$ can be freely chosen. }
    \label{fig:Switching signal}
\end{figure}

Furthermore, by carefully selecting the signals fed to the operators $\Emme_2$ and $\Emme_\infty$, it is possible to embed contextual information into the rPB controller without altering its properties, thereby aiding the computation of effective control actions. As shown in \citep{kirsch2025boostingtransientperformancereference}, one way to do so is by passing these sequences only through $\Emme_\infty$. Here, we propose an alternative approach that allows these signals to also pass through $\Emme_2$. To achieve this, we leverage the structure of $\mathbf{p}$. In particular, this sequence can be used to design a windowing signal that clamps all signals to zero under nominal conditions. Accordingly, we define the rectangular windowing signal $\boldsymbol{\sigma} \in \ell^1$ as:
\begin{equation*}
    \sigma_t = 
    \begin{cases}
    1 & \text{if  } || p_{t}||>\epsilon \\
    0 & \text{otherwise}
    \end{cases},
\end{equation*}
with $\epsilon \in \mathbb{R}$ a numerical tolerance. By using $\boldsymbol{\sigma}$ it is possible to window any signal $ \bs \in \ell^b$ as $s_{\sigma,t} = s_t*\sigma_t$. An intuitive way to understand is to see $\boldsymbol{\sigma}$ as a signal controlling a switching gate letting the other signals pass through to $\Emme_2$ or not. A block diagram of this setup can be found in Figure \ref{fig:Switching signal}. By construction, $\bs_{\sigma}$ is an  $\ell_p$ signal. Introducing the windowing signal enables non-$\ell_p$ contextual information to be fed to $\Emme_2$, improving its expressivity without compromising steady state guarantees. Incorporating this contextual information requires redefining $\Emme$ by adding a fourth argument. The operator now maps $\Emme : \ell^{8} \times \ell^{3} \times \ell^{3} \times \ell^{c} \to \ell^{m}$, where $c$ specifies the dimension of the contextual information.

\begin{remark}
Passing the contextual information through $\Emme_\infty$ ensures that it can influence the behavior of the rPB controller even when the switch in Figure \ref{fig:Switching signal} is open. 
\end{remark}

Up to this point, we have only used the fault occurrence window for defining the time interval during which the rPB mechanism remains active. However, the transition back to nominal operating conditions also induces a short transient phase, during which keeping the rPB active may be beneficial. This allows for a smoother transition from the fault period with rPB to the steady-state under the base controller alone. Maintaining the switching signal active for a short, finite duration after the fault can therefore improve the post-fault response. Importantly, extending the rPB activity window by a finite amount does not compromise the $\ell_p$ nature of the outputs of $\Emme$, and the steady-state guarantees are preserved.

\subsection{Interconnection with the base system}

As mentioned in Section \ref{rpb control}, the interconnection between the rPB output $\ub$ and the base system is unrestricted. In the base controller proposed in \citep{chandrasekaran2025reactivepowerflowoptimization}, the modulation vector $m_{g,t}$ is saturated at each timestep to enforce input constraints. We choose to make $\ub$ act as an offset sequence to the modulation vector before saturation. The modulation vector is now computed as:  
\begin{align*} 
        m_{g,t}&= \operatorname{sat}_{\|\cdot\|\leq \frac{1}{\sqrt{2}}} (m^b_{g,t}+u_t),
\end{align*}
where $m^b_{g,t}$ is the modulation vector coming from the base controller and here $u_t\in \mathbb{R}^2$. We choose this interconnection because it is the place where the influence of the rPB can be the strongest, while still preserving input constraints thanks to the saturation. 

By leveraging the windowing approach described in section \ref{signal}, we have considerable freedom in choosing the signals that we feed to $\Emme_2$. We choose $\bs$ such that for all $t$: 
\begin{equation*}
    s_t = [{v_{dc,t}},{i_{g,t}}^\top,{v_{g,t}}^\top ]^\top,  
\end{equation*}
which is then windowed using $\mathbb{\sigma}$ to obtain $\bs_\sigma$ (see Figure \ref{fig:Switching signal}). We also provide $\hat \bp$ to $\Emme_2$, but without windowing, as it is already an $\ell_2$ signal. We choose to feed $\bs_\infty = [\db^\top, \bp^\top, \boldsymbol{{v_{dc}}}, \boldsymbol{{i_g}}^\top]^\top$ to $\Emme_\infty$, as it conveys valuable information on the electrical state of the drive which is impacted by the faults. We thus get: 
\begin{equation*}
    \ub = \Emme_2(\bs_\sigma, \hat \bp)*\Emme_\infty(\bs_\infty). 
\end{equation*}
Furthermore, we feed these signals in per units. This serves as normalization, which facilitates the training for the neural network. Indeed, as the output of the rPB is directly saturated at $1/\sqrt{2}$, feeding it values orders of magnitude bigger than this bound would greatly hinder the learning.

\section{Simulation results} \label{Sec4}
This section describes the simulation results when applying the rPB approach to the drive for fault compensation. The main goal of the rPB controller is to preserve the DC bus voltage within a 2.5\% range of its nominal voltage $v_{dc}^\textit{ref}$, while also keeping the grid current under its maximum acceptable value.  The loss function \eqref{NOC2:cost} of the NOC problem is thus: 
\begin{align*}
    L(\eta_{T:0},u_{T:0}, y_{ref,T:0}) &= \sum_{t=0}^T \alpha_{nom}(v_{dc,t}-v_{dc}^\textit{ref})^2\\
    &+\alpha_{r,v_{dc}}b(v_{dc,t})+\alpha_{r,i_{g}}b(||i_{g,t}||),
\end{align*}

where the terms $\alpha_{nom}$, $\alpha_{r,v_{dc}}$, and $\alpha_{r,i_{g}}$ are scaling factors. The function $b:\mathbb{R}\rightarrow\mathbb{R}$ is a barrier function, which we choose as a squared ReLU: 
\[
b(x) = \text{ReLU}(x - x_{\max})^2 + \text{ReLU}(x_{\min} - x)^2,
\] 
where $x_{min}$ and $x_{max}$ are the bounds. We parametrize $\Emme_2$ as a REN with internal linear and nonlinear states of dimension twenty-two. For $\Emme_\infty$, we use a MLP consisting of four linear layers with widths 6, 10, 10, and 2, and apply sigmoid activations after each hidden layer.


\subsection{Single phase drop}

We consider nominal operation where the load on the mechanical side is $\tau_l = 0.95\tau_{max}$, where $\tau_{max}=44.356kNm$ is the maximum torque the drive can supply. The drive operates at nearly full capacity. The setpoints to track are $v_{dc}^\textit{ref}=5000V$, $w^\textit{ref}=125.66 rad/s$ and $Q^\textit{ref}=0VAR$. The constraints are $i_{g,max}= 2222A$ and $v_{dc}\in [0.975v_{dc}^\textit{ref},1.025v_{dc}^\textit{ref}]$.

We compare the response to a single phase loss fault with and without the rPB outer loop. We consider faults where the C phase of the grid voltage drops by a percentage value $\Delta \in \mathbb{R}$. Given the fault $t_{start}$ and $t_{end}$ times, the grid voltage in ABC coordinates is taken as:
\begin{align*}
    &v_{g,t}^A = |V_g|cos(2\pi ft)\\
    &v_{g,t}^B = |V_g|cos(2\pi ft-2\pi/3)\\
    &v_{g,t}^C = 
    \begin{cases}
    (1-\Delta)|V_g|cos(2\pi ft+2\pi/3) & \text{if  } t\in[t_{start},t_{end}] \\
    |V_g|cos(2\pi ft+2\pi/3) & \text{otherwise}
    \end{cases}
\end{align*}

where $|V_g|=3150V$ is the nominal grid voltage norm and $f=50Hz$ is the grid frequency.

To train the rPB controller, we generate a dataset of $v_g$ sequences containing drops of phase C of around $300$ms, and converted to $\alpha\beta$ coordinates. We slightly change the length of the fault between different profiles to cover endings occurring over the entire oscillation period of the grid voltage. The profiles also differs in the scale of the drop, with $\Delta$ sampled from a uniform distribution between $0$ and $1$. The generated dataset is composed of 300 samples. 

As outlined in Section~\ref{signal}, $\sigma$ is held at $1$ briefly after the fault is cleared. In this implementation, the switch remains closed until the next zero crossing of $v_{g,\beta}$. Enforcing the switching instant at a zero crossing results in a smoother transition, since the $\beta$-component coincides with its fault-free value at that point.

We train the rPB controller using stochastic gradient descent with Adam for 2600 epochs, setting a learning rate of $1\times 10^{-3}$.

\begin{figure}[h]
    \centering
    \includegraphics[width=\linewidth]{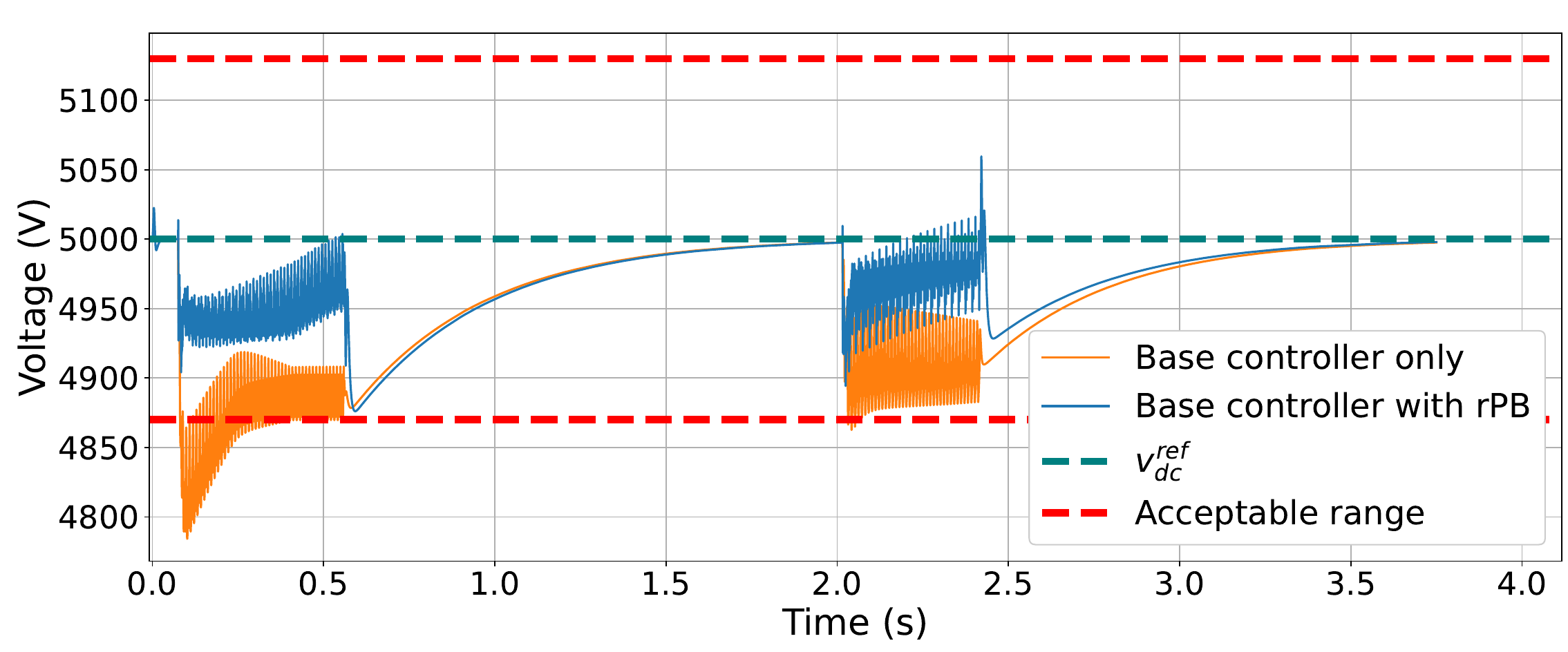}
    \caption{Evolution of the DC bus voltage with and without the trained rPB controller}
    \label{fig:vdc_profile}
\end{figure}

The $v_{dc}$ profiles with and without the rPB controller are shown in Figure \ref{fig:vdc_profile} for a test profile (not seen during control design) with two faults: the first one is a full drop of phase ($\Delta=1$) lasting for about 600ms and the second one is a $60\%$ drop lasting for around 400ms. Both event end at different times within the grid voltage period. Without rPB, the AFE controller struggles to keep the DC bus voltage within bounds, reaching a minimum below 4800 V, whereas with rPB the lower bound is never violated. Furthermore, the base controller induces constraints violations for more than 300ms during the first fault, which would be enough time to trigger a drive shutdown. By  allowing the system to stay within safe regions the rPB allows the continuation of operation even during faults, significantly improving the performance of the drive. During the second fault, the base controller mostly keeps the DC bus voltage within the bounds but strongly reduces it while the rPB controller keeps the DC bus voltage closer to its nominal value, reducing the strain on the drive, and showing that the neural network generalizes well to several depths of drops. Looking at the periods in between faults, we can see that the trajectories with and without rPB converge to the same trajectory tracking the nominal value, showcasing how the rPB controller does not hinder steady state properties.
\begin{figure}[h]
    \centering
    \includegraphics[width=\linewidth]{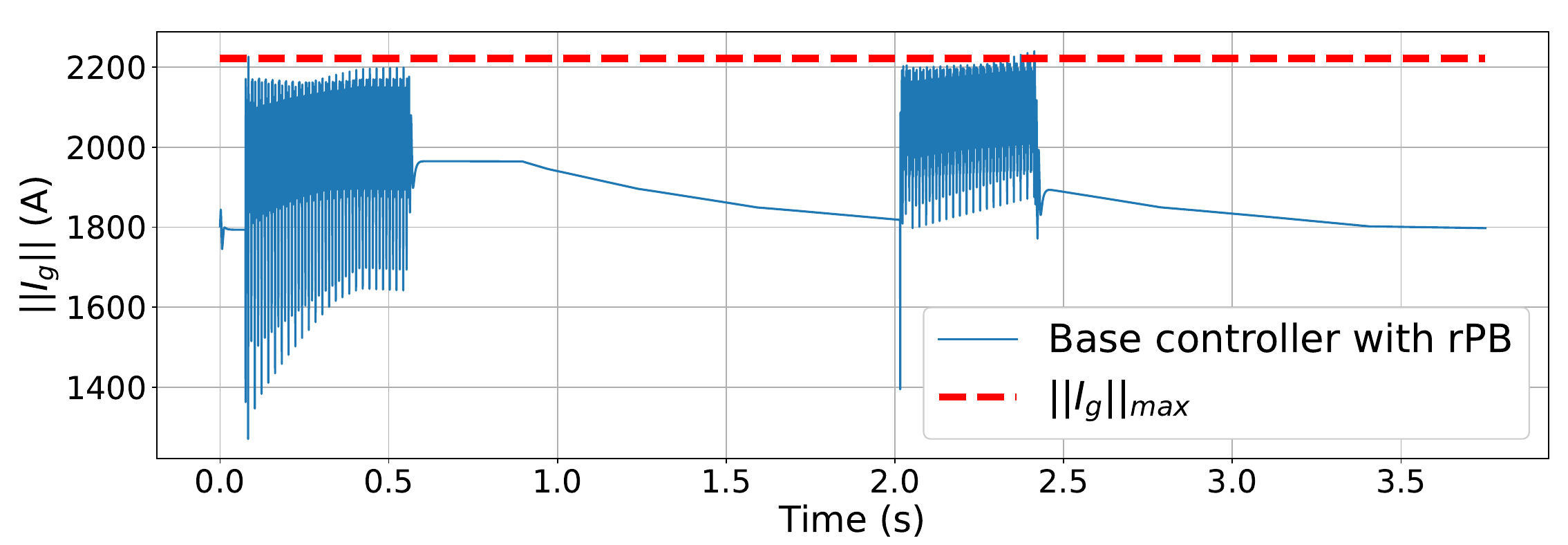}
    \caption{Evolution of the grid current norm with the trained rPB controller}
    \label{fig:ig_profile}
\end{figure}

Although the main objective for the rPB controller in this experiment is to keep the DC bus in safe operating conditions, it is important to see if it does so at the expense of the other states of the drive. Currently, rPB has no influence on the mechanical quantities of the drive as $\tau_m$ is not modified by it. rPB therefore cannot have a negative impact on the angular speed. For the grid side, Figure \ref{fig:ig_profile} shows the evolution of the grid current norm under the rPB outer loop. The neural network controller effectively learns to keep the current norm below its prescribed upper bound. Only brief, minor peaks slightly exceed the limit, and the rPB does not cause any sustained constraint violations. Such short-lived violations are acceptable for drive operation and are too small to pose any risk to the system. Overall, the rPB controller improves the DC bus voltage regulation during faults without compromising the safety of the other drive states.

\begin{figure}[h]
    \centering
    \begin{subfigure}[b]{0.49\linewidth}
        \centering
        \includegraphics[width = \linewidth]{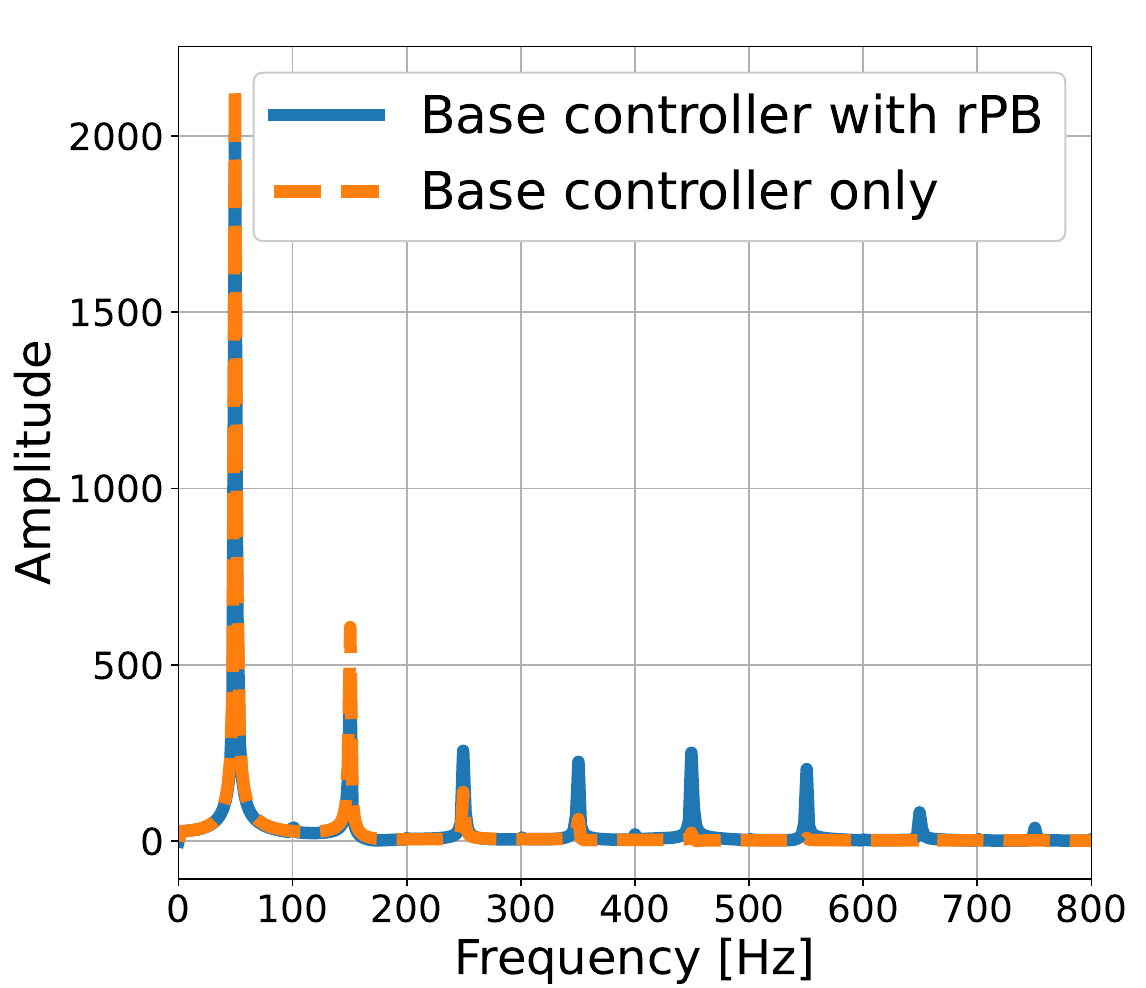}
        \caption{$\alpha$ coordinate}
        \label{fig:alpha}
    \end{subfigure}
    \begin{subfigure}[b]{0.49\linewidth}
        \centering
        \includegraphics[width = \linewidth]{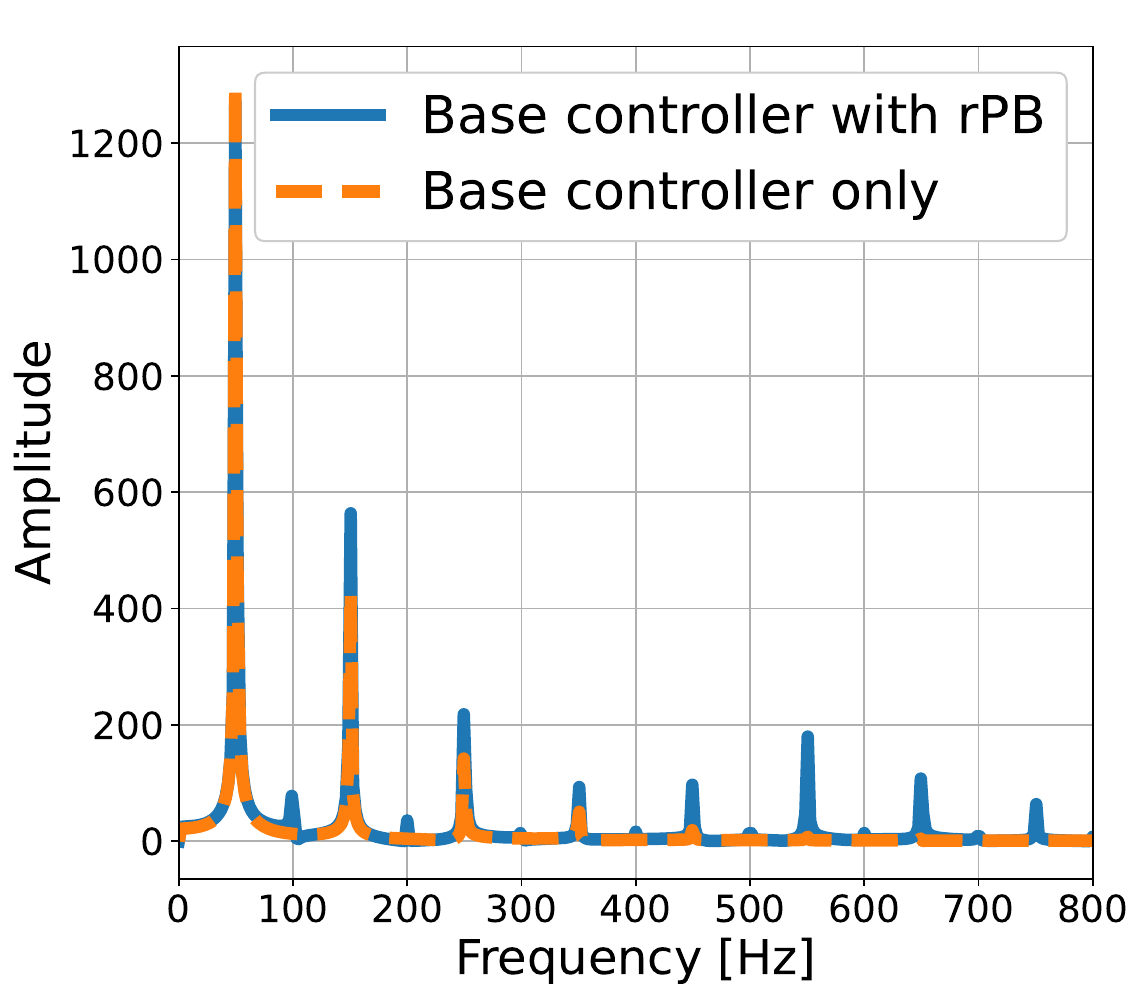}
        \caption{$\beta$ coordinate}
        \label{fig:beta}
    \end{subfigure}
    \caption{Frequency spectra of the $\alpha$ and $\beta$ coordinates of the grid current with and without the rPB controller.}
    \label{fig:alpha-beta spectra}
\end{figure}

Both with and without rPB, the faults induce oscillations in the DC bus voltage and in the grid current. It is therefore worth analyzing the spectra for both control strategies to assess whether some problematic harmonics have been generated, especially for the grid current. The spectra of the $\alpha$ and $\beta$ coordinates of the grid current during the first fault are shown in Figure \ref{fig:alpha-beta spectra}.

The main peak in both the $\alpha$ and $\beta$ spectra occurs at 50 Hz under both control schemes, corresponding to the grid’s fundamental frequency. The most significant harmonics appear at 150 Hz and 250 Hz due to the single-phase loss, which is typical for such grid faults. Both approaches exhibit similar magnitudes at these lower-order harmonics, indicating that the fundamental and main harmonic content is predominantly governed by the fault and not drastically altered by the rPB addition. For higher frequencies the controller with rPB introduces relatively smaller harmonics, which are most pronounced around 550 Hz and 650 Hz in the $\beta$ coordinate. These additional peaks remain significantly lower in amplitude than the one at 50 Hz. The emergence of these higher-order harmonics in the presence of rPB can be attributed to the rPB controller’s dynamics in response to unbalanced fault conditions. Nevertheless, their limited amplitude suggests that the influence of rPB is not a concern for the grid.

\begin{figure}
    \centering
    \includegraphics[width=\linewidth]{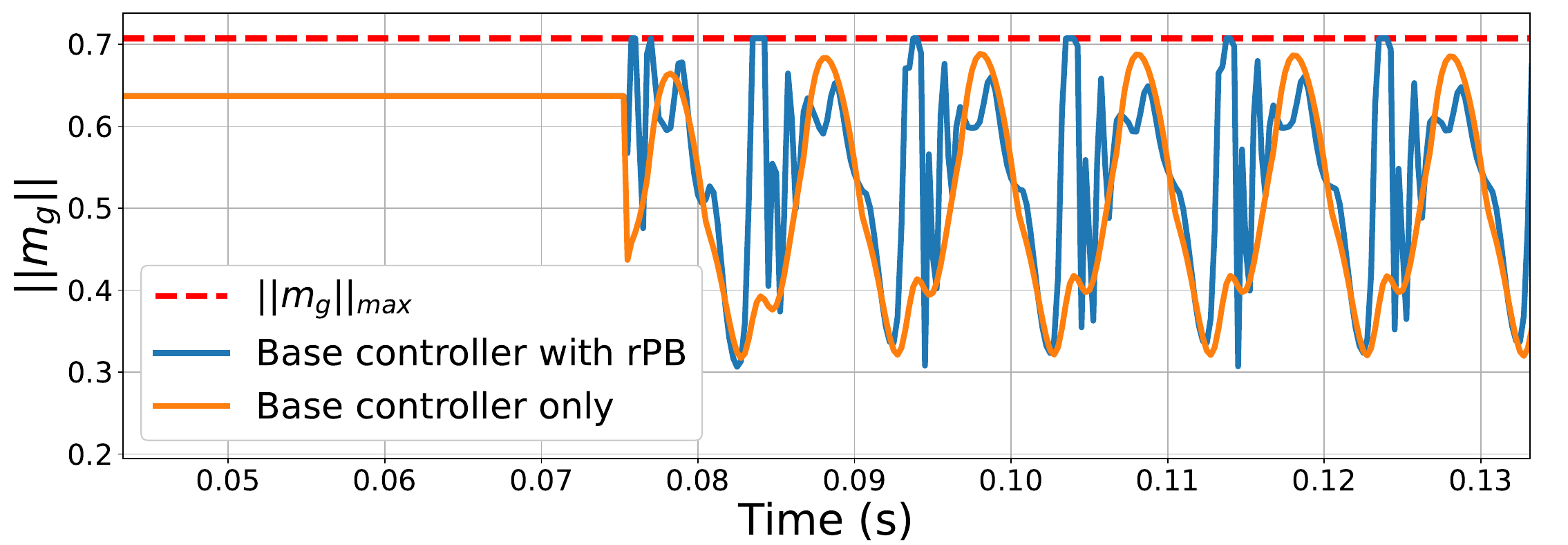}
    \caption{Evolution of the modulation vector norm with and without rPB at the end of the fault.}
    \label{fig:mg}
\end{figure}

Another relevant aspect to examine is how the rPB outer loop alters the modulation vector $m_g$. Figure \ref{fig:mg} shows the norm of the modulation vector around the beginning of the first fault. In both the baseline and rPB-modified cases, the signal is periodic during the fault. However, the profile resulting from the rPB controller exhibits a less regular intra-period structure. The overall norm is greater when utilizing rPB, which underscores its role in increasing the effective control strength. Focusing on the onset of the fault, the rPB strategy induces two distinct peaks where the modulation norm reaches its maximum permissible value, whereas the baseline controller's norm decreases over this interval. These peaks increase the instantaneous modulation amplitude and therefore allow the DC bus to draw more power from the grid during the fault, helping the DC bus to compensate the initial strong voltage drop. Because the rPB offset is introduced prior to actuator saturation, the input constraints remain satisfied over the whole trajectory.

In the interest of space, we have limited the results presented here to the electrical states of the drive, where the rPB has the most pronounced impact. More comprehensive simulation results are provided in Appendix \ref{Appendix}.

\section{Conclusion} \label{Sec5}

The rPB framework has been successfully integrated into a medium voltage drive system with an active front-end, ensuring safe and continuous operation during various grid fault scenarios. By designing input signals that leverage the grid voltage norm, rPB activation is precisely timed to coincide with the presence and duration of faults, independent of their recurrence or length. Our results with single-phase faults demonstrate that rPB reduces the DC bus voltage drop, supporting safe drive operation, and keeps the grid current within permissible limits. In the event of a phase loss near full load, the rPB controller prevents violations severe enough to otherwise trigger a system shutdown. These benefits underline the strong operational resilience provided by rPB. Future research should explore extending rPB's effect to the mechanical side of the drive, by enabling it to actuate the motor torque.  Another promising direction is to make the rPB able to address all classes of grid faults. Currently, rPB is tailored for a specific case, but incorporating more contextual inputs could enable fault-type-agnostic compensation. A possible way to do so would be to feed the prediction of of a dynamic fault classifier to the rPB controller.

\begin{figure*}[!t]
    \centering
    \includegraphics[width=\linewidth]{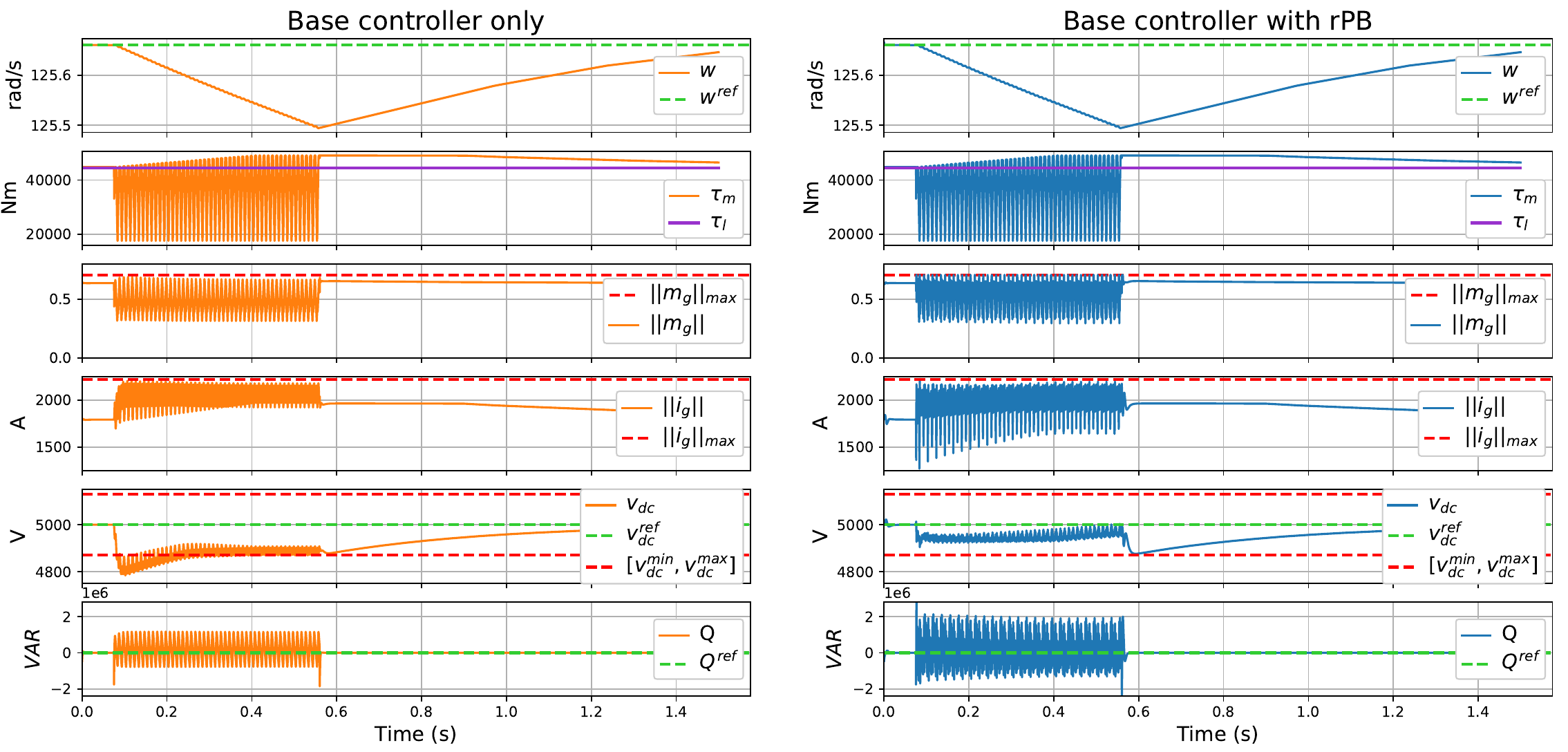}
    \caption{The first two rows show the behaviour of the mechanical variables. The third row illustrates the evolution of the modulation vector, and the last three rows present the electrical quantities.}
    \label{fig:all_together}
\end{figure*}

\bibliography{ifacconf}             
                                                   







\appendix
\section{Complementary results}  \label{Appendix}  
We present in here the complementary results to the ones presented in the main paper. The evolution of the key drive parameters for the first $1.5$ seconds are plotted in Figure \ref{fig:all_together}. This interval contains the full loss of phase fault and the recovery towards nominal conditions. During the fault, we can see that the angular speed of the shaft decreases. This is because the drop in supplied grid voltage limits the maximum active power which can be delivered to the shaft and thus the torque. The available torque thus become insufficient to compensate the load, reducing the angular speed. Still, the scale of the drop is negligible, reaching at most $0.1\%$ of the nominal angular speed. An important observation is that the angular velocity  and the torque profiles are the same with and without rPB, which makes sense as it only influences the modulation vector $m_g$.                                                                          
\end{document}